\documentclass[12pt,letterpaper]{article}        
\usepackage[utf8]{inputenc}
\usepackage{jheppub}
\usepackage{epsfig}
\usepackage{amssymb,amsfonts}
\usepackage{amsmath}
\usepackage{changes}

\newcommand{\be} {\begin{equation}}
\newcommand{\ee} {\end{equation}}
\newcommand{\bea} {\begin{eqnarray}}
\newcommand{\eea} {\end{eqnarray}}

\title{Incoherent conductivity of holographic charge density waves}

\author[1]{Blaise Gout\'{e}raux,} 
\author[2]{Niko Jokela,} 
\author[2]{and Arttu Pönni}

\affiliation[1]{Nordita, KTH Royal Institute of Technology and Stockholm University, Roslagstullsbacken 23, SE-106 91 Stockholm, Sweden}

\affiliation[2]{Department of Physics and Helsinki Institute of Physics, P.O.Box 64, FIN-00014 University of Helsinki, Finland}

\emailAdd{blaise.gouteraux@su.se}
\emailAdd{niko.jokela@helsinki.fi}
\emailAdd{arttu.ponni@helsinki.fi}

\abstract{The DC resistivity of charge density waves weakly-pinned by disorder is controlled by diffusive, incoherent processes rather than slow momentum relaxation. The corresponding incoherent conductivity can be computed in the limit of zero disorder. We compute this transport coefficient in holographic spatially modulated breaking translations spontaneously. As a by-product of our analysis, we clarify how the boundary heat current is obtained from a conserved bulk current, defined as a suitable generalization of the Iyer-Wald Noether current of the appropriate Killing vector.}

\begin{document}

\begin{flushright}
HIP-2018-15/TH\\
NORDITA 2018-017
\end{flushright}

\maketitle

\pdfoutput=1
\pagestyle{plain} 
\setcounter{page}{2}
\newcounter{bean}

\baselineskip16pt

%
%
%
%
%
%
%
%
%
%


\section{Introduction} 

Phases breaking translations spontaneously play a prominent role in the phase diagram of strongly-correlated Condensed Matter systems, such as high $T_c$ superconductors. After being anticipated on theoretical grounds \cite{PhysRevB.40.7391,MACHIDA1989192,MACHIDA19901047}, they were subsequently observed experimentally \cite{Tranquada:1995}.
In holography, spatially modulated instabilities of translation invariant phases have been thoroughly studied, see e.g. \cite{Nakamura:2009tf,Donos:2011bh,Bergman:2011rf}. The corresponding backreacted, spatially modulated phases have been constructed as well \cite{Donos:2013wia,Withers:2013loa,Withers:2014sja,Jokela:2014dba,Donos:2015eew,Jokela:2016xuy,Cremonini:2016rbd,Cremonini:2017usb,Cai:2017qdz} and are dual to various kinds of strongly-coupled density waves.

As translations are not explicitly broken, momentum is still conserved and the DC conductivities are formally infinite \cite{RevModPhys.60.1129,Delacretaz:2016ivq,Delacretaz:2017zxd}
\begin{equation}
\label{ACcond}
\sigma(\omega)=\sigma_o+\frac{\rho^2}{\chi_{PP}}\left(\frac{i}\omega+\pi\delta(\omega)\right).
\end{equation}
In the formula above, $\rho$ is the charge density of the state and $\chi_{PP}=\delta P/\delta v$ the momentum static susceptibility. $\sigma_o$ is a transport coefficient that appears at first order in gradients in the constitutive relation of the current density
\begin{equation}
j=\rho v-\sigma_o\partial\mu+\dots
\end{equation}
with $\mu$ the chemical potential and $v$ the velocity. The dots stand for terms unimportant to the conductivity calculation. At zero density and without broken translations, $\sigma_o$ would represent the quantum critical conductivity due to particle-hole pair creation in the vacuum \cite{Herzog:2007ij}. At non-zero density, it captures the contribution of incoherent, diffusive processes which do not drag momentum, \cite{Davison:2015bea,Davison:2015taa}.

It can be defined more formally by a Kubo formula \cite{Davison:2015taa}
\begin{equation}
\label{eq:incconddef}
\sigma_o=\frac1{\chi_{PP}{}^2}\lim_{\omega\to0}\frac{\textrm{Im}G^R_{J_{inc}J_{inc}}(\omega,q=0)}{\omega}\ .
\end{equation}
 It involves the incoherent current 
 \begin{equation}
 \label{eq:inccurrentdef}
 J_{inc}=\chi_{PP}J-\rho P\ ,
 \end{equation}
 which by construction is orthogonal to momentum, $\chi_{J_{inc}P}=0$. $\sigma_o$ has been computed holographically in translation-invariant phases \cite{Hartnoll:2007ip,Jain:2010ip,Davison:2015taa}, phases with weak momentum relaxation \cite{Davison:2015bea} as well as phases with spontaneous translation symmetry breaking \cite{Amoretti:2017frz,Amoretti:2017axe}. In the latter case, the breaking was realized homogeneously. The purpose of this note is to generalize this computation to inhomogeneous, spatially modulated black branes which break translations spontaneously.
 
For simplicity, we focus on a parity-preserving Einstein-Maxwell-dilaton model, \eqref{eq:action}, where \cite{Donos:2013gda} has shown spatially modulated instabilities arise given certain conditions on the behavior of the scalar couplings in the infra red. We will restrict to spontaneous breaking in one spatial direction only. Our starting point will be the general construction of \cite{Donos:2014yya}, turning on an external electric field and a temperature gradient at the boundary. What we will show, as noticed in \cite{Davison:2018}, is that for spontaneous boundary conditions in the UV, requesting certain metric elements to fall off sufficiently fast at the boundary imposes a specific relation between the electric field and the temperature gradient. This is equivalent to a rotation of sources, which itself implies that only the incoherent current is sourced and not momentum.
 
One novelty of our setup is the presence of a pure gauge solution to the equations of motion, which can be obtained by acting on the static background with a Lie derivative along the spatially modulated direction. This can loosely be thought of as the Goldstone mode of spontaneous translation symmetry breaking, the phonon. This mode contributes to the local, spatially dependent currents and consequently to the local incoherent conductivity. As we shall see, it drops out after spatial averaging over the system, and so does not appear in the zero mode of the ac conductivity \eqref{ACcond}. It can be interpreted as the sliding velocity of the density wave and cannot be fixed simply from data at the horizon, as pointed out in \cite{Jokela:2016xuy}.

Another technical point we clarify is how to define properly the boundary heat current from a conserved current in the bulk. How this works out for the spatial component of the heat current has been extensively studied in past holographic literature, starting with \cite{Donos:2014cya}. Drawing on \cite{Kastor:2008xb,Kastor:2009wy}, we show that a conserved bulk current can be defined such that its time component asymptotes to the time component of the boundary heat current. The main technical concept is based on a generalization of the Iyer-Wald Noether charge \cite{Wald:1993nt,Iyer:1994ys} involving Killing potentials.\footnote{Connections between the spatial component of the holographic heat current and the Iyer-Wald formalism were noted previously in \cite{Liu:2017kml}.} This leads us to an improved definition \eqref{eq:defQM} of the heat current compared to holographic literature, which turns out to be crucial to properly understand the effect of a non-zero sliding velocity on the spatial currents.
 
 \paragraph{Note added:\\} As this work was nearing completion, \cite{Donos:2018kkm} appeared which contains some overlap with our results.


\section{Background} \label{sec:background}

In this paper we study a family of actions in a $(3+1)$-dimensional bulk spacetime. Our starting point is the Einstein-Maxwell-dilaton action, which reads as follows
\be\label{eq:action}
   S = \frac{1}{16\pi G_N}\int d^4 x \sqrt{-g} \left( R - \frac{1}{2}\partial\phi^2 - \frac{Z(\phi)}{4}F^2 - V(\phi) \right) \ ,
\ee
where the functions $Z$ and $V$ only depend on the scalar $\phi$ and are left unspecified for the time being. The equations of motion following from (\ref{eq:action}) are
\bea
 R_{\mu\nu} +\frac{1}{2}Z F_{MS}F^S_{\ N}-\frac{1}{2}\partial_M\phi\partial_N\phi & = & \frac{1}{2}g_{MN}\left( R - \frac{1}{2}\partial\phi^2 - \frac{Z}{4}F^2 - V \right)  \\ \label{eq:ELg}
 \frac{1}{\sqrt{-g}}\partial_M\left(\sqrt{-g} \partial^M\phi\right) & = & \frac{1}{4}Z' F^2 +V' \\  \label{eq:ELphi}
 \frac{1}{\sqrt{-g}}\partial_M\left(\sqrt{-g} Z(\phi) F^{MN}\right) & = & 0 \ .\label{eq:ELF}
\eea

We will focus on asymptotically locally AdS$_4$ solutions to (\ref{eq:ELg})-(\ref{eq:ELF}) which have a regular Killing event horizon in the IR and exhibit spontaneous translation symmetry breaking in one of the field theory directions that we take to be $x$. To this end we adopt the following 
Ansatz \cite{Donos:2014yya}
\bea
   g    & = & -U(r) H_{tt}(r,x) dt^2 + \frac{H_{rr}(r,x)}{U(r)} dr^2 + \Sigma(r,x) \left( e^{B(r,x)} dx^2 + e^{-B(r,x)} dy^2 \right) \label{eq:ansatz_g} \\
   A    & = & a_t(r,x) dt \label{eq:ansatz_A} \\
   \phi & = & \phi(r,x) \label{eq:ansatz_phi} \ ,
\eea
where our convention for the radial coordinate $r$ is such that the boundary resides at $r=\infty$ and the horizon is at $r=r_h$. We furthermore restrict to the case where all the functions are periodic in $x$ with period $L$, except $U$ which only depends on $r$ without loss of generality.

Restricting to asymptotically AdS$_4$ solutions as $r\to\infty$ imposes several conditions on the scalar functions and the solutions themselves. First of all, the scalar vanishes, leading to the following expansions
\be
   V(\phi\to 0)  = -6 + \frac{1}{2} m^2 \phi^2 + \ldots \ , \quad   Z(\phi\to 0)  =  1 + \ldots \ ,
\ee
where we take the scalar mass $m^2=-2$ in the following to simplify our boundary expansions. Second, the Ans\"atze for the metric components, Maxwell potential $a_t$, and the scalar $\phi$ are expanded as follows
\be
\label{eq:backUVexp}
\begin{split}
   U(r)  = &  r^2 + \frac{U^{(\infty)}}{r} + \ldots \\
   H_{tt}(r,x) = & 1 + \frac{H_{tt}^{(\infty)}(x)}{r^3} + \dots \\
   H_{rr}(r,x)  = & 1 + \frac{H_{rr}^{(\infty)}(x)}{r^3} + \dots \\
   \Sigma(r,x)  = & r^2 + \frac{\Sigma^{(\infty)}(x)}{r} + \dots \\
   B(r,x) = & \frac{B^{(\infty)}(x)}{r^3} + \dots \\
   a_t(r,x)   = & \mu - \frac{\rho(x)}{r} +\ldots \\
   \phi(r,x)  = & \frac{\phi^\text{vev}(x)}{r^2}+\ldots \ ,
   \end{split}
\ee
compatible with AdS$_4$ asymptotics. To ensure translations are only broken spontaneously, none of the source terms including the chemical potential $\mu$ depend on $x$, while the vevs $\rho(x)$ and $\phi^\text{vev}(x)$ are generically $x$-dependent functions. We also note that the boundary metric is simply the Minkowski metric.

Regularity at the horizon $r\to r_h$ yields the following expansion
\be
\label{eq:backreghor}
\begin{split}
   U(r\to r_h)  = & 4\pi T (r-r_h) + \ldots \\
   H_{tt}(r\to r_h)  = & H_{rr}(r\to r_h) = H_{tt}^{(0)}(x) + \ldots \\
   \Sigma(r\to r_h)  = & \Sigma^{(0)}(x) + \ldots \\
   e^{B(r\to r_h)} = & e^{B^{(0)}(x)} + \ldots \\
   a_t(r\to r_h)  =  & (r - r_h) a_t^{(0)}(x) + \ldots \\
   \phi(r\to r_h)  = & \phi^{(0)}(x) + \ldots \ ,
   \end{split}
\ee
where $\ldots$ represent terms that vanish faster as $r\to r_h$.
This can be checked by changing to ingoing Eddington-Finkelstein (EF) coordinates, which read close to the horizon
\be
t\mapsto v-\frac1{4\pi T}\ln(r-r_h)+\ldots \ .
\ee
$T$ is the usual Hawking temperature, which can be computed by requiring that the periodicity $\beta=1/(2\pi T)$ of imaginary time of the Euclidean solution is such that there is no conical singularity at $r=r_h$.

The behavior of the gauge field also follows from going to EF coordinates and requiring $a_M dx^M$ be regular at $r=r_h$.

\subsection{Pure gauge solution and sliding velocity \label{section:gaugemode}}

The background solution is not unique. The following linearized coordinate transformation\footnote{It would be interesting to generalize the following discussion to non-linear Lorentz boosts, but for our purposes it is enough to work to linear order in $\delta v_s$.}
\be
\label{eq:coordchangevs}
t\mapsto t-\delta v_s x\,,\qquad x\mapsto x-\delta v_s t
\ee
together with the gauge transformation $A\mapsto A+d\Lambda$, $\Lambda(x)=\delta v_s \mu x$ also yields solutions solving the background equations to linear order in $\delta v_s$, after a suitable modification of the horizon regularity conditions \eqref{eq:backreghor}. Note that it is crucial to perform all of these operations simultaneously to avoid introducing new sources at the boundary. $\delta v_s$ is a constant which is not fixed by the background equations. Physically, it represents the freedom for the CDW to slide and is directly connected to the existence of a Goldstone mode due to spontaneous translation symmetry breaking \cite{Jokela:2016xuy}. At the level of the background solution, it is consistent to pick the gauge $v_s=0$ where the background is time-independent. This is the gauge we work with for simplicity in the remainder of this work.

\subsection{Boundary stress-energy tensor}

To extract the energy-momentum tensor from the bulk metric, we go to the Fefferman-Graham gauge by means of the following change of radial coordinate:
\be
\label{FGcoordTrafo}
   \frac{dz}{z} = \sqrt{\frac{H_{rr}}{U}} dr \ .
\ee
In the Fefferman-Graham coordinates the metric close to the boundary $z\to 0$ expands as
\be
   g = \frac{1}{z^2} (dz^2 + dx_\mu dx^\mu + z^3 H_{\mu\nu}(z,x) dx^\mu dx^\nu + \dots ) \label{eq:ef_metric} \ .
\ee
The boundary energy-momentum tensor $\langle T_{\mu\nu}(x) \rangle = \frac{3}{16\pi G_N} H_{\mu\nu}(z=0,x)$ can then be directly read off from the metric and is spatially dependent.
After performing the coordinate transformation \eqref{FGcoordTrafo}, we end up with a metric in the form (\ref{eq:ef_metric}) from which we extract the stress-energy tensor
\bea\label{eq:Tmunu}
    T  & = & -(2 U^{(\infty)}+ H_{rr}^{(\infty)} + 3 H_{tt}^{(\infty)}) dt^2 + (-U^{(\infty)}+3 B^{(\infty)}+H_{rr}^{(\infty)}+3\Sigma^{(\infty)}) dx^2 \nonumber \\ 
  & &  + (-U^{(\infty)}-3 B^{(\infty)}+H_{rr}^{(\infty)}+3\Sigma^{(\infty)}) dy^2 \ .
\eea
Here, and in the rest of the article, we have set $16\pi G_N=1$. 
By further using the UV expansions of the metric functions, we find the following constraints
\bea
   H_{tt}^{(\infty)} + H_{rr}^{(\infty)} + 2\Sigma^{(\infty)} & = & 0 \\
   (H_{tt}^{(\infty)} - \Sigma^{(\infty)} - 3 B^{(\infty)})' & = & 0 \ ,
\eea
which are the dilatation and diffeomorphism Ward identities obeyed by the stress tensor
\bea
    T_\mu^\mu  & = & 0 \\
   \partial^\mu  T_{\mu\nu}  &  = & \partial_x  T_{xx}  = 0 \ .
\eea
These are not the only constraints on the boundary data. As we are considering phases breaking translations spontaneously, we should also require that the free energy is minimized with respect to the periodicity, which identifies the preferred spatially modulated phase. We will return to this at the end of the next section.

\subsection{Charge and entropy density from Noether currents}

The solutions we are after have two conserved quantities that will be of interest in the following. The first one is the total charge density on the boundary corresponding to the global $U(1)$ gauge symmetry (see eg \cite{Papadimitriou:2005ii}). 
The gauge field equation of motion (\ref{eq:ELF}) states that the bulk current 
\be
\label{eq:elcurrentdef}
\mathcal J^{M}=\sqrt{-g}Z(\phi)F^{M r}
\ee
is conserved
\be
\label{U1conscurrent}
   \nabla_M\mathcal  J^M = \frac1{\sqrt{-g}}\partial_M (\sqrt{-g} Z F^{M r}) = 0 \ .
\ee
In our ansatz only the temporal component of the field strength is non-zero, implying the following radially conserved current
\bea
   \partial_r (\sqrt{-g} Z F^{rt}) + \partial_x (\sqrt{-g} Z F^{xt}) & = & 0 \\
   \rightarrow \partial_r \left( \int \sqrt{-g} Z F^{rt} \right) & = & 0 \ .
\eea
Here we have adopted the notation $\int := L^{-1} \int_0^L dx$ for the spatial averaging. Since the above is radially conserved, we can directly evaluate $\mathcal J^t$ at the boundary and link it with the average charge density of the operator dual to $A$,
\be
 \bar J^t=  \int \sqrt{-g} Z F^{tr}\Bigg|_{r=\infty} = \int \frac{Z(\phi) \Sigma(r,x) \partial_r a_t(r,x)}{\sqrt{H_{tt}(r,x) H_{rr}(r,x)}}\Bigg|_{r=\infty} = \int \rho(x) \equiv \bar{\rho} \ .
\ee

The other conserved quantity is related to the entropy density and requires slightly more work to write in closed form. We will eventually find a radially conserved current that evaluates to $sT$. Our starting point is the antisymmetric two-form \cite{Papadimitriou:2005ii,Donos:2014cya,Donos:2014yya}
\be
   G^{MN} = \nabla^M k^N + \frac{1}{2} Z k^{[M}F^{N]I} A_I + \frac{1}{4}(\psi-2\theta) F^{MN} \ ,
\ee
where $k=\partial_t$ is a Killing vector of our solution  ($\mathcal{L}_k g = \mathcal{L}_k F = \mathcal{L}_k \phi = 0$). The functions $\psi$ and $\theta$ are solutions to $L_k A=d\psi$ and $i_k F=d\theta$. At the practical level, this means that $\psi=0$ and $\theta=-a_t$. Furthermore, $G^{MN}$ satisfies
\be
\label{conseqG}
   \nabla_N G^{MN} = - \frac{V(\phi)}{2} k^N \ .
\ee
This is not quite on the same footing as the electric current \eqref{U1conscurrent}, as $G^{MN}$ is not conserved. This can be remedied by the following argument, connected to the so-called Noether entropy current and Komar integrals in the General Relativity literature \cite{Wald:1993nt,Iyer:1994ys,Kastor:2008xb,Kastor:2009wy}. \cite{Kastor:2008xb,Kastor:2009wy} in particular were concerned with gravity with a cosmological constant ($A=\phi=0$). We recall their arguments here, and then will generalize them to the case at hand. Since $k$ is a Killing vector, it obeys the Killing equation and so is divergenceless $\nabla_M k^M=0$. This immediately implies that it can locally be expressed in terms of an antisymmetric two-form $k^M=\nabla_N \omega^{NM}$. This relation is not unique, as we can always shift $\omega^{MN}$ by a co-closed antisymmetric 2-form $\lambda^{MN}$, $\tilde \omega^{MN}=\omega^{MN}+\lambda^{MN}$. A simple choice is
\be
\label{eq:lambdadef}
\lambda^{rt}=-\lambda^{tr}=\frac{\alpha}{\sqrt{-g}}
\ee 
and all other components zero, which can readily be checked to verify $\nabla_M\lambda^{MN}=0$ at background level. This implies is does not affect \eqref{conseqG}, which can be rewritten
\be
\label{conseqGLambda}
   \nabla_M G^{MN} = - \Lambda \nabla_M \tilde\omega^{MN}\quad\Leftrightarrow\quad     \nabla_M\left( G^{MN} +\Lambda \tilde\omega^{MN}\right)=0\ .
\ee
The improved bulk current $G^{MN} +\Lambda \tilde\omega^{MN}$ is now manifestly conserved.  We will see shortly that its $rt$-component gives the heat density $sT$ after a suitable choice of $\alpha$ in \eqref{eq:lambdadef}. It also makes it clear that both currents ultimately originate from bulk symmetries.

Now let us go through the same steps in the case with a non-zero scalar field. We observe that 
\be
\nabla_M\left(V(\phi)k^M\right)=V(\phi)\nabla_M k^M+V'k^M\nabla_M\phi=0
\ee
where the first term vanishes because $k$ is a Killing vector, and the second because $\mathcal L_k\phi=0$. Thus, we expect we should be able to find a two-form such that $V(\phi)k^M=\nabla_N \omega^{NM}$. Indeed, we find by direct computation that\footnote{This fact was first noticed in the context of a different collaboration between B.G., Richard Davison and Simon Gentle involving translation-invariant black hole solutions to Einstein-Maxwell-dilaton theories \cite{Davison:2018}.}
\bea
\label{nontrivialstep}
   \partial_M (\sqrt{-g} G^{M t}) + \sqrt{-g}\frac{V(\phi)}{2} = \partial_r (\sqrt{-g} G^{rt}) + \partial_x (\sqrt{-g} G^{xt}) \nonumber \\
   + \frac{1}{2} \partial_r (\sqrt{-g} \omega^{rt}(r,x)) + \frac{1}{2} \partial_x (\sqrt{-g} \omega^{xt}(r,x)) = 0 \ ,
\eea
where $\omega^{rt}$ and $\omega^{xt}$ are functions involving only metric functions and their derivatives (see Appendix~\ref{sec:appendix} for details of the derivation). Similarly as above, this yields a radially conserved current
\be\label{ConseqGrt}
   \partial_r \int \left( \sqrt{-g} G^{rt} + \frac{1}{2} \sqrt{-g} \omega^{rt}+\frac\alpha2 \right) = 0 \ .
\ee
Setting $\alpha$ to zero for now, we can then evaluate \eqref{ConseqGrt} at the horizon to show that it is related to the entropy density $s$:
\bea \label{sT}
   \int  \sqrt{-g}\left( G^{rt} + \frac{1}{2} \omega^{rt} \right)\Bigg|_{r=r_h} & = & \int \frac{\Sigma^2}{4\sqrt{-g}} \left( \Sigma\partial_r\left(\frac{U H_{tt}}{\Sigma}\right) - Z a_t \partial_r a_t + U H_{tt} \partial_r B \right)\Bigg|_{r=r_h} \nonumber \\ 
   & = & \pi T \int \Sigma^{(0)}(x) = \frac{1}{4} s T \ ,
\eea
where in the last step we identified the average Bekenstein-Hawking entropy density $s=\frac{1}{4}\int \Sigma^{(0)}(x)$ with the entropy density of the boundary theory. Anticipating on the analysis of linear fluctuation and to have the correct normalization for the spatial components of the heat current, we now make the gauge choice 
\be
\alpha=\frac{sT}{2}
\ee
so that in the end we define the bulk heat current
\be
\label{eq:defQM}
\mathcal Q^M=2 \sqrt{-g} G^{rM} +\sqrt{-g}\tilde\omega^{rM}\,.
\ee
Its zero mode is radially conserved
\be
   \partial_r \int\mathcal Q^t(r,x) = 0 \
\ee
and
\be \label{eq:sTfinal}
   \int \mathcal Q^t(r,x)\Bigg|_{r=r_h}= s T \ .
\ee
This makes it clear that the entropy density (times temperature) is the Noether charge associated to the timelike Killing vector $k=\partial_t$ \cite{Papadimitriou:2005ii}. 
  Since the current is radially conserved, we can also evaluate it at the boundary:
\begin{align}\label{GrtUV}
   \int \mathcal Q^t\Bigg|_{r=\infty} &= \frac{1}{2} \int \left(  T_{tt}(x) +  T_{yy}(x) - \mu \rho(x) \right)+\frac{sT}{2} \ .
\end{align}
We note that it is crucial to take into account the second term $\omega^{rt}$ in order to renormalize the boundary divergence contained in $G^{rt}$.

Putting together \eqref{sT} and \eqref{GrtUV} returns an integral Komar (Smarr) relation:
\be
\label{Komar}
sT+\mu\bar J^t=\bar T^{tt}+\bar T^{yy}\ .
\ee
We noted above an ambiguity in the definition of $\omega^{MN}\to\omega^{MN}+\lambda^{MN}$. We see that this ambiguity does not affect the integral relation we have just derived: since $\lambda$ is itself closed, its contributions at the boundary and at the horizon are of equal magnitude but opposite sign, and so drop out from \eqref{Komar}.

In \cite{Donos:2015eew} it was shown that the free energy density for this class of theories read $w=-sT-\mu \bar J^t+\bar T^{tt}$, and that moreover minimizing it with respect to the periodicity (to find the most stable phase) implied the condition $w+\bar T^{xx}=0$. Thus we deduce that in fact $\bar T^{xx}=\bar T^{yy}$ and $B^{(\infty)}=0$ from \eqref{eq:Tmunu}. We further obtain
\be
\bar T^{xx}=\bar T^{yy}=p=\frac12\bar T^{tt}=\frac12\bar\epsilon\,,\quad sT+\mu\bar J^t=\bar\epsilon+p\ ,
\ee
which gives a Smarr-type relation for the background thermodynamic quantities. Thanks to the underlying relativistic structure of the boundary theory, we can boost this stress-energy tensor using a velocity $u^\mu$ to
\be
\bar T^{\mu\nu}=(p+\bar\epsilon)u^\mu u^\nu+p \eta^{\mu\nu} \ .
\ee
From there, we can compute the momentum static susceptibility by linearizing the averaged stress-energy tensor around the equilibrium solution
\be
\chi_{PP}=\frac{\delta \bar T^{tx}}{\delta v^x}=\bar\epsilon+p
\ee
which matches the result in \cite{Donos:2018kkm}.


\section{The incoherent conductivity}

In this section we determine the thermoelectric DC conductivities of our system in terms of horizon data.

\subsection{Perturbation ansatz}
With a straightforward generalization from \cite{Jokela:2016xuy} we turn on the following perturbations
\begin{align}
   g & \mapsto g + \left(\delta g_{tt}+\delta v_s t U(r)\partial_x H_{tt}\right) dt^2 + 2 \delta g_{tr} dtdr+ \left(\delta g_{rr}-\delta v_s t \frac{\partial_x H_{rr}}{U(r)}\right) dr^2   \nonumber \\
   &+ 2 \left(\delta g_{tx} -\xi H_{tt} U t+\delta v_s H_{tt}U(r)-\delta v_s \Sigma e^B\right) dtdx+ 2 \delta g_{rx}dxdr  \nonumber\\
   &+\left(\delta g_{xx}-\delta v_s t \partial_x(\Sigma e^B)\right) dx^2 + \left(\delta g_{yy}-\delta v_s t \partial_x(\Sigma e^{-B})\right) dy^2 \\
   A & \mapsto A + \left(\delta a_t-\delta v_s t \partial_x a_t\right) dt + \delta a_r dr + \left(\delta a_x + a_t \xi t - E t -\delta v_s a_t+\delta v_s\mu\right) dx \\
   \phi & \mapsto \phi + \delta\phi - \delta v_s t \partial\phi \ .
\end{align}
$E$ is a constant and uniform electric field which sources the electric current, $\xi$ a constant and uniform temperature gradient which sources the heat current.\footnote{These sources can also be made periodic \cite{Donos:2015gia}.} $E$ and $\xi$ appear such that perturbation equations of motion are time independent when background functions are on-shell.\footnote{Setting $\delta v_s=0$, the time dependence introduced by $E$ and $\xi$ can be removed by a coordinate transformation $t\mapsto t(1-\zeta\, x)$ and gauge transformation $A\mapsto A+d\Lambda$, $\Lambda=t\,E\,x$, \cite{Donos:2015bxe}.} $\delta v_s$ terms can be generated through similar gauge and coordinate transformations as in section \ref{section:gaugemode}. In contrast to the background, they cannot be gauged away since we have now turned on sources linear in $t$, on which the coordinate transformation \eqref{eq:coordchangevs} would act. Indeed, \cite{Jokela:2016xuy} found such terms were necessary to match the AC and DC computation of the electric conductivity in a probe brane setup. All other perturbations are assumed to be periodic in $x$, and decay sufficiently fast at the boundary not to introduce any other source.

Horizon regularity imposes additional constraints on perturbations, which we have collected in Appendix \ref{app:horizon}.

\subsection{Currents and conductivity}

Taking our cue from \cite{Donos:2014yya}, we look for two conserved bulk currents that asymptote to the spatial component of the electric and heat currents, respectively. We first focus on the electric current. Since the CDW slides $\delta v_s\neq0$, $\mathcal J^x=\sqrt{-g} Z(\phi) F^{x r}$ is no longer conserved but instead it is a function depending on $r$ and $x$. Indeed: 
\bea
   \partial_t ( \sqrt{-g} Z(\phi) F^{tr} ) + \partial_x (\sqrt{-g} Z(\phi) F^{xr} ) & = & 0 \\
   \partial_t ( \sqrt{-g} Z(\phi) F^{tx} ) + \partial_r (\sqrt{-g} Z(\phi) F^{rx} ) & = & 0 \ .
\eea
Non-zero temporal derivatives spoil the conservation of $\mathcal J^x$, so we must find a new combination that is conserved.
This combination is found by observing that to first order, the following holds
\begin{align}
   \partial_t ( \sqrt{-g} Z(\phi) F^{tr} ) &= -\delta v_s \partial_x ( \sqrt{-g} Z(\phi) F^{tr} ) \\
   \partial_t ( \sqrt{-g} Z(\phi) F^{tx} ) &= -\delta v_s \partial_x ( \sqrt{-g} Z(\phi) F^{tx} ) \ .
\end{align}
These together with equations of motion imply that
\begin{align}
   \partial_r (\mathcal  J^x - \delta v_s\mathcal  J^t ) &= 0 \\
   \partial_x (\mathcal  J^x -\delta  v_s\mathcal  J^t ) &= 0 \ .
\end{align}
Thus, we find the following conserved quantity analogous to electric current
\be\label{eq:tildeJ}
 \tilde{\mathcal J}^x := \mathcal J^x -\delta v_s \mathcal J^t \ .
\ee
We observe that $\mathcal J^M$ defined in \eqref{eq:elcurrentdef} transforms under the coordinate change \eqref{eq:coordchangevs} with $v_s\mapsto\delta v_s$ in such a way to exactly compensate the second term in \eqref{eq:tildeJ}. So \eqref{eq:tildeJ} is the combination invariant under \eqref{eq:coordchangevs}. Indeed we can check by direct computation that 
\be 
\label{eq:bulkJxtrafo}
\mathcal J^x = \mathcal J^x(\delta v_s=0)+\delta v_s\mathcal  J^t\,.
\ee
So all $\delta v_s$ dependence drops out from $\tilde{\mathcal J^x}$.

This of course has a natural interpretation. As the translation symmetry breaking is assumed to be spontaneous, the CDW does not have a preferred location to reside. This does not impede constructing such inhomogeneous solutions numerically by picking an origin of the $x$-coordinate and forcing the solution, for example, to have a zero phase there. When one turns on a constant, uniform electric field perturbation $E$, in absence of impurities or pinning potentials, the CDW will immediately react due to Lorentz force and begin sliding. The traveling CDW carries with itself the charge carriers and the natural conserved current one would write down is (\ref{eq:tildeJ}).

We also note that \eqref{eq:bulkJxtrafo} is consistent with how any boundary current $J^\mu=(J^t,\delta J^x,0)$ transforms under \eqref{eq:coordchangevs}:
\be
\label{eq:bdyJxtrafo}
J^\mu\to J^\mu=(J^t-t\delta v_s \partial_x J^t,\delta J^x+\delta v_s J^t,0) \ .
\ee

A similar story holds for the heat current, with a subtlety related to the two-form $\omega^{MN}$. To first order in perturbations, the Killing vector $k$ ($\mathcal{L}_k g = \mathcal{L}_k F = \mathcal{L}_k \phi = 0$) is
\begin{align}
   k = (1-\xi x)\partial_t + \delta v_s \partial_x \ ,
\end{align}
implying
\begin{align}
   \theta &= -(1-\xi x)a_t(r,x) - \delta a_t(r,x) - E x + \delta v_s t \partial_x a_t(r,x) \\
   \psi &= -E x  \ .
\end{align}
With these choices for $k$, $\theta$ and $\psi$ we know that the two-form satisfies \eqref{conseqG}. We need the $r$ and $x$-components which are more explicitly
\begin{align}
   \partial_t ( \sqrt{-g} G^{tr} ) + \partial_x ( \sqrt{-g} G^{xr} ) &= 0 \\
   \partial_t ( \sqrt{-g} G^{tx} ) + \partial_r ( \sqrt{-g} G^{rx} ) &= \delta \left(\frac{-\sqrt{-g} V(\phi)}{2} k^x \right) \ .
\end{align}
First notice that at background level, $k^x=0$ and at first order $\delta k^x=\delta v_s$, meaning that up to first order
\begin{align}
   - \frac{\sqrt{-g} V(\phi)}{2} k^x = - \delta v_s \frac{\sqrt{-g} V(\phi)}{2} \ .
\end{align}
Again the following holds
\begin{align}
   \partial_t ( \sqrt{-g} G^{tr} ) &= -\delta v_s \partial_x ( \sqrt{-g} G^{tr} ) \\
   \partial_t ( \sqrt{-g} G^{tx} ) &= -\delta v_s \partial_x ( \sqrt{-g} G^{tx} )
\end{align}
which brings the conservation equations of $G^{\mu\nu}$ to the following form
\bea
   \partial_x \left( \sqrt{-g} G^{xr} - \delta v_s \sqrt{-g} G^{tr} \right) & = & 0 \\
   \partial_r ( \sqrt{-g} G^{rx} ) - \delta v_s \partial_x ( \sqrt{-g} G^{tx} ) & = & - \delta v_s \frac{\sqrt{-g}V(\phi)}{2} \ .
\eea
The second equation can we rewritten, using background equations for $G^{MN}$ as
\begin{gather}
   \partial_r ( \sqrt{-g} G^{rx} +\delta  v_s \sqrt{-g} G^{tr} ) -\delta  v_s \frac{\sqrt{-g}V(\phi)}{2} = -\delta v_s \frac{\sqrt{-g}V(\phi)}{2} \ ,\\
  \Rightarrow \partial_r ( \sqrt{-g} G^{rx} +\delta  v_s \sqrt{-g} G^{tr} )=0 \ .
\end{gather}
The same observations can be made about the combination $\sqrt{-g} G^{rx} +\delta  v_s \sqrt{-g} G^{tr} $ as for $\tilde{\mathcal J^x}$: this is a combination invariant under \eqref{eq:coordchangevs}, it has no $\delta v_s$ dependence left.

However, the combination $\sqrt{-g} G^{rx} +\delta  v_s \sqrt{-g} G^{tr} $ does not match the expected  transformation of the boundary heat current \eqref{eq:bdyJxtrafo}.
Indeed, since $\int 2\sqrt{-g} G^{rx}$ asymptotes to the zero mode of the heat current when $\delta v_s=0$, we would have expected it to transform as 
\be
2\sqrt{-g} G^{rx}\mapsto 2\sqrt{-g} G^{rx}+\delta v_s \mathcal Q^t \ ,
\ee
where $\mathcal Q^t$ was defined in \eqref{eq:defQM}. This discrepancy comes precisely from taking into account the contribution of the-two form $\tilde\omega^{MN}$. From the previous equations, we know it verifies $\nabla_M\tilde \omega^{Mx}=0$, otherwise it would have contributed explicitly. But it should also transform appropriately under \eqref{eq:coordchangevs}:
\begin{equation}
\delta\tilde \omega^{rx}=-\delta v_s \tilde\omega^{rt}=-\delta v_s \left(\omega^{rt}+\frac{sT}{2\sqrt{-g}}\right)\,.
\end{equation}
Combined with how $G^{rx}$ is expected to transform, this indeed gives us the correct combination $\mathcal Q^t$.

At the end of the day, we define
\be
\label{eq:tildeQ}
 \tilde{\mathcal Q}^x = 2\sqrt{-g}\left(G^{rx} - \delta v_s G^{rt}\right) = \mathcal Q^x - \delta v_s \mathcal Q^t \ ,\quad \mathcal Q^x=2\sqrt{-g}G^{rx}  -\delta v_s \left(\sqrt{-g}\omega^{rt}+\frac{sT}2\right)
\ee
in analogy to (\ref{eq:tildeJ}). As for the electric current, this amended heat current is conserved $\partial_r \tilde{\mathcal Q}^x = \partial_x \tilde{\mathcal Q}^x = 0$ and finite at the boundary.

Now that we have the conserved quantities $\mathcal{J}^x$ and $\mathcal{Q}^x$, we proceed to evaluate them on the black hole horizon and extract the associated horizon conductivities. At leading order we obtain
\begin{align}
   \tilde{\mathcal J}^x_{(0)} &= e^{-B^{(0)}(x)} Z(\phi^{(0)}(x))\left(E + \partial_x\delta a_t^{(0)}(x)\right) \nonumber \\
   &\hspace{3cm} - \frac{Z(\phi^{(0)}(x)) a_t^{(0)}(x)}{H_{tt}^{(0)}(x)} \left(\delta g_{tx}^{(0)}(x)+\delta v_s \Sigma^{(0)}(x)\right) \\
   \tilde{\mathcal Q}^x_{(0)} &= -4\pi T \left(\delta g_{tx}^{(0)}(x)+\delta v_s \Sigma^{(0)}(x)\right) 
\end{align}
which verify
\be
\nabla_x \tilde{\mathcal J}^x_{(0)}=0\ ,\qquad\nabla_x \tilde{\mathcal Q}^x_{(0)}=0\ .
\ee
By expanding $\tilde{Q}^x$ to next order in $r-r_h$ we find an additional equation
\begin{gather}
   \partial_x \left( 4\pi T \frac{\delta g_{tr}^{(0)}(x)}{H_{tt}^{(0)}(x)} - \frac{\delta g_{tx}^{(0)}(x)+\delta v_s \Sigma^{(0)}(x)}{\Sigma^{(0)}(x)} \partial_x \left(B^{(0)}(x) - \log(H_{tt}^{(0)}(x) \Sigma^{(0)}(x))\right) \right) \nonumber \\
    + \frac{\delta g_{tx}^{(0)}(x)+\delta v_s \Sigma^{(0)}(x)}{\Sigma^{(0)}(x)} \left( \partial_x \log \frac{e^{B^{(0)}(x)}}{\Sigma^{(0)}(x)} \right)^2 + \frac{(\partial_x \phi^{(0)}(x))^2}{\Sigma^{(0)}(x)} \left(\delta g_{tx}^{(0)}(x)+\delta v_s\Sigma^{(0)}(x)\right) \nonumber \\
   + \frac{Z(\phi^{(0)}(x)) a_t^{(0)}(x)}{H_{tt}^{(0)}(x)} (E + \partial_x\delta a_t^{(0)}(x))+ 4\pi T \xi = 0 \ .
\end{gather}
Notice that even though any explicit $\delta v_s$ dependence had dropped out from $\tilde{\mathcal J}^x$, $\tilde{\mathcal Q}^x$, it reappears in the equations above due to the horizon regularity conditions.

These can be used to solve for $\tilde{\mathcal{J}}_x$ and $\tilde{\mathcal{Q}}_x$ in terms of background functions at $r=r_h$. After some algebra, we obtain 
\be
\begin{split}
\label{eq:tilde_E_xi}
 \tilde{\mathcal{J}}^x=& {\sigma}_h E + \alpha_h \xi  \\
   \tilde{\mathcal{Q}}^x =& {\bar{\alpha}}_h E + \bar{\kappa}_h \xi  \ ,
   \end{split}
\ee
where
\begin{align}
   {\sigma}_h = & \frac{\int \{\dots\}}{\int \frac{e^{B^{(0)}(x)}}{Z(\phi^{(0)}(x))} \int \{ \dots \} - \left( \int \frac{e^{B^{(0)}(x)} a_t^{(0)}(x)}{H_{tt}^{(0)}(x)} \right)^2 } \\
   \alpha_h = {\bar{\alpha}}_h =& \frac{4\pi T \int \frac{e^{B^{(0)}(x)} a_t^{(0)}(x)}{H_{tt}^{(0)}(x)}}{\int \frac{e^{B^{(0)}(x)}}{Z(\phi^{(0)}(x))} \int \{ \dots \} - \left( \int \frac{e^{B^{(0)}(x)} a_t^{(0)}(x)}{H_{tt}^{(0)}(x)} \right)^2 } \\
   \bar{\kappa}_h =& \frac{(4\pi T)^2 \int \frac{e^{B^{(0)}(x)}}{Z(\phi^{(0)}(x))}}{\int \frac{e^{B^{(0)}(x)}}{Z(\phi^{(0)}(x))} \int \{ \dots \} - \left( \int \frac{e^{B^{(0)}(x)} a_t^{(0)}(x)}{H_{tt}^{(0)}(x)} \right)^2 } \\
   \int \{\dots\} = & \int \Bigg\{ \frac{e^{B^{(0)}(x)} Z(\phi^{(0)}(x)) a_t^{(0)}(x)^2}{H_{tt}^{(0)}(x)^2} \nonumber \\
   & + \frac{1}{\Sigma^{(0)}(x)} \left( \partial_x \log \frac{e^{B^{(0)}(x)}}{\Sigma^{(0)}(x)} \right)^2 + \frac{(\partial_x \phi^{(0)}(x))^2}{\Sigma^{(0)}(x)} \Bigg\} \ .
\end{align}
Notice that all of these transport coefficients are guaranteed to be positive by a Schwarz-inequality
\begin{align}
   \int \frac{e^{B^{(0)}(x)}}{Z(\phi^{(0)}(x))} \int \frac{e^{B^{(0)}(x)} Z(\phi^{(0)}(x)) a_t^{(0)}(x)^2}{H_{tt}^{(0)}(x)^2} \geq \left( \int \frac{e^{B^{(0)}(x)} a_t^{(0)}(x)}{H_{tt}^{(0)}(x)} \right) \ .
\end{align}
Here we emphasize that these horizon conductivities have no meaning by themselves in the boundary theory. There, since translations are not broken explicitly, all physical conductivities diverge as $\omega\to0$, see \eqref{ACcond}.

The quantity which is physical at the boundary is captured by the incoherent conductivity \eqref{eq:incconddef}. As explained in the introduction, it is given by a Kubo formula involving the boundary incoherent current \eqref{eq:inccurrentdef}. We can then write down a bulk current that asymptotes to it:
\begin{align}
\label{IncCurrent}
   \mathcal J_{inc}^x(r,x) := sT\mathcal  J^x(r,x)-\bar\rho \mathcal Q^x(r,x)\ .
\end{align}

Actually, the equations of motion and UV boundary conditions force us to consider this particular combination. Requiring $\delta g_{rx}$ to fall-off sufficiently fast at the boundary, the $rx$-component of metric perturbation equations near the boundary implies
\begin{align}
   (E - \mu \xi)\bar \rho + \xi \left(\bar  T^{tt} + \bar T^{xx} \right)= 0 \ .
\end{align}
This can be simplified using $sT = \bar \epsilon + p - \mu\bar \rho = \bar T^{tt}+ \bar T^{xx}- \mu\bar \rho$. We end up with
\begin{align}
   -E = \frac{sT}{\bar \rho} \xi =- sT \alpha_{inc} \ .
\end{align}
This is equivalent to rotating the sources from $(E,\xi)$ to $(\alpha_{inc},0)$. We are then led to rotating the currents $J^x$, $Q^x$, and find that $\alpha_{inc}$ is the source for the incoherent current, which is given by the linear combination of the original currents in \eqref{eq:inccurrentdef}. 

Plugging \eqref{eq:tildeJ} and \eqref{eq:tildeQ} in \eqref{IncCurrent}, we obtain
\be
   \mathcal J_{inc}^x(r,x)=  \tilde{\mathcal J}_{inc}^x(r,x)+\delta v_s\left(sT\mathcal J^t-\bar\rho\mathcal Q^t\right).
\ee
From the conservation of $\mathcal J^x$ and $\mathcal Q^x$, the zero mode of the incoherent current is radially conserved, and moreover
\be
\bar{\mathcal J}_{inc}^x = \int\tilde{\mathcal J}_{inc}^x \ .
\ee
Evaluating it at the horizon and using \eqref{eq:tilde_E_xi}, we can read off the spatially averaged incoherent conductivity
\be
\label{eq:sigmainchorizon}
\sigma_{inc}=\frac{\bar{\mathcal J}_{inc}^x}{\alpha_{inc}}=(sT)^2 \sigma_h-2sT\bar\rho\alpha_h+\bar\rho^2 \bar{\kappa}_h
\ee
This is related to the finite contribution to the real part of the AC conductivity through \eqref{eq:incconddef},\be
\label{eq:sigmainchorizon}
\sigma_o=\frac{\sigma_{inc}}{\chi_{PP}{}^2}\ .
\ee

\section{Discussion and outlook}

In real systems, translations are inevitably broken explicitly as well, for instance by disorder or inelastic scattering of the charge carriers with the underlying lattice. If translations are weakly broken, the long wavelength effective theory of clean charge density waves is modified in two ways. Firstly, momentum relaxes slowly, which is captured by introducing a momentum relaxation rate $\Gamma$. Secondly, the Goldstone mode (the phonon) acquires a small mass, but can remain light enough that it does not decouple from the dynamics. 

The AC conductivity at low frequencies becomes
\begin{equation}
\label{accondpinnedCDW}
\sigma(\omega)=\sigma_o+\frac{(\chi_{JP})^2}{\chi_{PP}}\frac{-i\omega}{-i\omega(\Gamma-i\omega)+\omega_o^2}\,.
\end{equation}
$\omega_o$ is the pinning frequency, which is directly proportional to the phonon mass. The AC conductivity of a pinned CDW looks quite different from that of a weakly-disordered metal: it has a finite frequency peak at $\omega=\omega_o$ rather than a Drude-like peak centered at $\omega=0$. The DC resistivity is no longer controlled by momentum relaxation. Indeed, setting $\omega=0$ in \eqref{accondpinnedCDW} returns
\begin{equation}
\rho_{dc}=\frac1{\sigma_o}+O(\Gamma,\omega_o)\,.
\end{equation}
The resistivity is no longer small as in a metal with slow momentum relaxation, where $\rho_{dc}\sim O(\Gamma)$. Instead, it is governed by the incoherent conductivity $\sigma_o$. As $\sigma_o$ is insensitive to momentum dynamics at leading order, it can be computed in the clean state without disorder. This is precisely the computation we have carried out in this work, and what our formula \eqref{eq:sigmainchorizon} captures. The interplay between weak disorder and the Goldstone dynamics short-circuits the effects of momentum relaxation on the DC resistivity and this generally leads to bad metallic behavior, \cite{Delacretaz:2016ivq}.

Pinned collective modes of phases with spontaneous symmetry breaking have been reported in previous holographic literature \cite{Jokela:2017ltu,Alberte:2017cch,Andrade:2017cnc}. In particular, \cite{Jokela:2017ltu,Andrade:2017ghg} computed the  resistivity of an inhomogeneous spatially modulated phase. Both of the setups contain a term violating parity, and it would be interesting to generalize our results in this direction, starting from \cite{Donos:2017mhp}. In their case, the phase is insulating at low temperatures. It would also be worthwhile to connect to the proposal of \cite{Delacretaz:2016ivq} by realizing `metallic' CDW phases, with a resistivity decreasing at low temperatures. 

\medskip

\acknowledgments
We thank Richard Davison, Matti J\"arvinen, Alexander Krikun, and Matthew Lippert for discussions. We especially thank Matti J\"arvinen for comments on a previous version of this manuscript. 
BG has been partially supported during this work by the Marie Curie International Outgoing Fellowship nr 624054 within the 7th European Community Framework Programme FP7/2007-2013. N.~J. and A.~P. have been supported in part by the Academy of Finland grant no. 1303622. A.~P. also acknowledges support from the Magnus Ehrnrooth foundation.

\appendix

\section{Derivation of $sT$}\label{sec:appendix}

The non-trivial step we skipped in Section \ref{sec:background} is the derivation of the equality
\begin{align}
   \sqrt{-g} V(\phi) = \partial_r (\sqrt{-g} \omega^{rt}) + \partial_x (\sqrt{-g} \omega^{rx}) \ .
\end{align}
This can be shown for general $V(\phi)$ by solving \eqref{eq:ELg} and assuming our ansatz \eqref{eq:ansatz_g}-\eqref{eq:ansatz_phi}. The $(x,x)$ and $(y,y)$-components of Einstein equations can be used to algebraically solve for $V(\phi)$
\begin{gather}
   V(\phi) = -\frac{U \partial_r B \partial_r H_{rr}}{4 H_{rr}^2}+\frac{e^{-B} \partial_x B \partial_x H_{rr}}{4 H_{rr} \Sigma}+\frac{U \partial_r B \partial_r H_{tt}}{4 H_{rr} H_{tt}}+\frac{\partial_r U \partial_r B}{2 H_{rr}}+\frac{U \partial_r B \partial_r \Sigma}{2 H_{rr} \Sigma }+\frac{U \partial_r^2 B}{2 H_{rr}} \nonumber \\
   +\frac{3 e^{-B} \partial_x B \partial_x H_{tt}}{4 H_{tt} \Sigma }+\frac{e^{-B} \partial_x B \partial_x \Sigma}{2 \Sigma^2}-\frac{e^{-B} (\partial_x B)^2}{2 \Sigma}+\frac{e^{-B} \partial_x^2 B}{2 \Sigma}-\frac{e^{-B} \partial_x H_{rr} \partial_x H_{tt}}{4 H_{rr} H_{tt} \Sigma}-\frac{e^{-B} \partial_x H_{rr} \partial_x \Sigma}{4 H_{rr} \Sigma^2} \nonumber \\
   -\frac{e^{-B} \partial_x H_{tt} \partial_x \Sigma}{4 H_{tt} \Sigma^2}+\frac{e^{-B} (\partial_x H_{tt})^2}{4 H_{tt}^2 \Sigma }-\frac{e^{-B} \partial_x^2 H_{tt}}{2 H_{tt} \Sigma}+\frac{e^{-B} (\partial_x \Sigma)^2}{2 \Sigma ^3}-\frac{e^{-B} \partial_x^2 \Sigma}{2 \Sigma ^2}+\frac{U \partial_r H_{rr} \partial_r H_{tt}}{4 H_{rr}^2 H_{tt}} \nonumber \\
   +\frac{\partial_r U \partial_r H_{rr}}{4 H_{rr}^2}+\frac{U \partial_r H_{rr} \partial_r \Sigma}{4 H_{rr}^2 \Sigma}-\frac{3 \partial_r U \partial_r H_{tt}}{4 H_{rr} H_{tt}}-\frac{3 U \partial_r H_{tt} \partial_r \Sigma}{4 H_{rr} H_{tt} \Sigma }+\frac{U (\partial_r H_{tt})^2}{4 H_{rr} H_{tt}^2}-\frac{U \partial_r^2 H_{tt}}{2 H_{rr} H_{tt}} \nonumber \\
   -\frac{\partial_r U'}{2 H_{rr}}-\frac{\partial_r U \partial_r \Sigma}{H_{rr} \Sigma }-\frac{U \partial_r^2 \Sigma}{2 H_{rr} \Sigma}.
\end{gather}
Notice that all terms have derivatives in them and only of one kind. With the Leibniz rule we can in the end of the day write the above as
\begin{align}
   \sqrt{-g} V(\phi) = \partial_r \underbrace{\left( \frac{\Sigma^2 U H_{tt} \partial_r B - \Sigma \partial_r(U H_{tt}\Sigma)}{2\sqrt{-g}} \right) }_{:= \sqrt{-g}\omega^{rt}}- \partial_x \underbrace{\left( \frac{H_{rr}\partial_r(H_{tt}\Sigma e^{B})}{2\sqrt{-g}} \right) }_{:=-\sqrt{-g}\omega^{xt}}.
\end{align}
Even though $\sqrt{-g}V(\phi)$ is not radially conserved, $\int \sqrt{-g}V(\phi)$ is since the integration makes the last term to vanish since all the metric functions are periodic in $x$. This fact is used in Section \ref{sec:background} to find a radially conserved quantity which asymptotes to $sT$.

\section{Near-horizon perturbations} \label{app:horizon}

The perturbations are required to be regular at the black hole horizon $r=r_h$. Regularity is ensured after switching to the ingoing Eddington-Finkelstein coordinate $v=t+(4\pi T)^{-1}\log(r-r_h)$ by expanding the perturbations in the following way at $r\to r_h$:
\begin{align}
   \delta g_{tt} &= U(r)\delta g_{tt}^{(0)}(x) + \delta v_s \frac{\log(r-r_h)}{4\pi T} U(r)\partial_x H_{tt}(r,x) \\
   \delta g_{rr} &= \frac{\delta g_{rr}^{(0)}(x)}{U(r)} - \delta v_s \frac{\log(r-r_h)}{4\pi T} \frac{\partial_x H_{rr}(r,x)}{U(r)} \\
   \delta g_{xx} &= \delta g_{xx}^{(0)}(x) - \delta v_s \frac{\log(r-r_h)}{4\pi T} \partial_x \left( \Sigma(r,x) e^{B(r,x)} \right) \\
   \delta g_{yy} &= \delta g_{yy}^{(0)}(x) - \delta v_s \frac{\log(r-r_h)}{4\pi T} \partial_x \left( \Sigma(r,x) e^{-B(r,x)} \right)  \\
   \delta g_{tr} &= \delta g_{tr}^{(0)}(x) \\
   \delta g_{tx} &= e^{B^{(0)}(x)} (\delta g_{tx}^{(0)}(x) + \delta g_{tx}^{(l)}(x) U(r) \log U(r)) + \delta v_s \Sigma(r,x) e^{B(r,x)} \\
   \delta g_{rx} &= \frac{e^{B^{(0)}(x)}}{U(r)} \delta g_{rx}^{(0)}(x) \\
   \delta a_t &= \delta a_t^{(0)}(x) -\delta v_s \frac{\log(r-r_h)}{4\pi T} \partial_x a_t(r,x) \\
   \delta a_r &= \frac{\delta a_r^{(0)}(x)}{U(r)} \\
   \delta a_x &= \log(r-r_h) (E - \xi a_t(r,x)) \delta a_x^{(0)}(x)
\end{align}
subject to
\begin{gather}
   \delta g_{rx}^{(0)}(x)-\delta g_{tx}^{(0)}(x) = 0, \quad \delta g_{tt}^{(0)}(x) + \delta g_{rr}^{(0)}(x) - 2 \delta g_{tr}^{(0)}(x) = 0, \\
   \delta a_r^{(0)}(x) - \delta a_t^{(0)}(x) = 0, \quad \delta a_x^{(0)}(x) = - \frac{1}{4\pi T}, \quad \delta g_{tx}^{(l)} = -\frac{e^{-B^{(0)}(x)}}{4\pi T} H_{tt}^{(0)}(x) \xi \ .
\end{gather}
\bibliographystyle{JHEP}
\bibliography{incoherentCDW-biblio}

\end{document}